\documentclass[aps,prl,twocolumn,showpacs,superscriptaddress]{revtex4-2}

\usepackage{amsmath}
\usepackage{graphicx}
\usepackage{lmodern}
\usepackage{amsmath}

\usepackage{color}
\usepackage{amssymb}
\usepackage{bm}
\usepackage{braket}
\usepackage{mathtools}
\newcommand{\bk}{\bm{k}}
\newcommand{\br}{\bm{r}}

\newcommand{\bq}{\bm{q}}

\DeclareMathOperator{\sgn}{sgn}

\newcommand{\blue}[1]{{\color{blue}{#1}}}

\usepackage[dvipsnames]{xcolor}
\usepackage[colorlinks=true]{hyperref}
\hypersetup{
    colorlinks=true,
    linkcolor=blue,
    citecolor=blue,
    urlcolor=blue
} 

\usepackage[normalem]{ulem}

\makeatletter
\def\maketitle{
\@author@finish
\title@column\titleblock@produce
\suppressfloats[t]}
\makeatother

\begin{document}

\title {Time-Reversal Invariant Topological Moiré Flatband: A Platform for the Fractional Quantum Spin Hall Effect }

\author{Yi-Ming Wu}
\affiliation{Stanford Institute for Theoretical Physics, Stanford
  University, Stanford, California 94305, USA}
\author{Daniel Shaffer}
\affiliation{Department of Physics, Emory University, 400 Dowman Drive, Atlanta, GA 30322, USA}
\author{Zhengzhi Wu}
\affiliation{Institute for Advanced Study, Tsinghua University, Beijing 100084, China}
\author{Luiz H. Santos}
\affiliation{Department of Physics, Emory University, 400 Dowman Drive, Atlanta, GA 30322, USA}

\date{\today}

\begin{abstract}
Motivated by recent observation of the quantum spin Hall effect in monolayer germanene and twisted bilayer transition-metal-dichalcogenides (TMDs), we study the topological phases of moiré twisted bilayers with time-reversal symmetry and spin $s_z$ conservation. 
By using a continuum model description which can be  applied to both germanene and TMD bilayers, we show that
at small twist angles, the emergent moiré flatbands can be topologically nontrivial due to inversion symmetry breaking.
Each of these flatbands for each spin projection admits a lowest-Landau-level description in the chiral limit and at magic twist angle. This allows for the construction of a many-body Laughlin state with time-reversal symmetry which can be stabilized by a short-range pseudopotential, and therefore serves as an ideal platform for realizing the so-far elusive fractional quantum spin Hall effect with emergent spin-1/2 U(1) symmetry.  

  
\end{abstract}
\maketitle

{\it Introduction.}-- 
Since the discovery of superconducting and correlated insulating phases in magic angle twisted bilayer graphene (TBG)\cite{Cao2018b,Xie2019,Jiang2019,Choi2019,Kerelsky2019,Matthew2019,Lu2019,Arora2020,Aaron2019,Li2020,ZhangYu2020,Saito2021}, the moiré engineering of 2D Van der Waals materials,  
{such as}
graphene 
{and}
transition metal dichalcogenides (TMDs)\cite{Tang2020,Regan2020,Xu2020,Jin2021,Huang2021,Li2021,Wang2020,Ghiotto2021,Zhang2020,Shabani2021,Weston2020}, 
{in a host of}
bilayer and multilayer {hetero}structures\cite{Liu2020,Shen2020,Chen2019a,Chen2019b,Chen2020,Cao2020b,He2021,Park2021,Zhu2020}, has attracted enormous research attention. 
{Moir\'e} platforms are ideal hubs 
{to forge}
the interplay between strongly correlated effects and topological phases, giving rise to 
{rich}
phenomena including superconductivity\cite{Cao2018b,Matthew2019,Lu2019,Arora2020}, strange metal behavior\cite{Polshyn2019,Cao2020a,Lyustrange}, magnetic quantum anomalous Hall effect in TBG\cite{Serlin2020,Aaron2019,Tseng2022} and more recently, 
{long-sought after fractional Chern insulators (FCI)}\blue{\cite{Neupert-2011,Sheng-2011,regnault2011fractional}}
first observed in twisted bilayer MoTe$_2$\cite{Cai2023,zeng2023integer,Park_2023,xu2023observation}.

{The discovery of FCIs in TMD heterostructures highlights the importance of moir\'e flat bands \cite{BM,PhysRevLett.99.256802,PhysRevLett.122.106405,Tarnopolsky2019,Wufc2018,Kang2018,Bultinck:2020,PhysRevB.103.205413} in the service of electron fractionalization without a magnetic field. In particular, moir\'e flatbands in the chiral limit \cite{Tarnopolsky2019} share properties akin to lowest Landau level (LLL) wavefunctions, shedding light on the stability of time-reversal broken topological states through local interactions\cite{PhysRevResearch.2.023237,PhysRevLett.124.046403,PhysRevLett.124.166601,PhysRevB.103.035427,PhysRevResearch.1.033126,PhysRevB.99.155415}. 
Conversely, a burning question arises: can moir\'e flat bands support \textit{time-reversal symmetric (TRS) fractional} topological order?
While TRS $\mathbb{Z}_2$ flatbands have been theoretically proposed in twisted bilayer TMDs\cite{Wufc2019,Pan2020,Zhang2021,PhysRevLett.130.126001,MaoKim23}, and experimental signatures of the quantum spin Hall (QSH) effect \cite{PhysRevLett.95.226801,PhysRevLett.95.146802,PhysRevLett.96.106802,doi:10.1126/science.1133734,PhysRevLett.97.036808,PhysRevB.90.081102,Goerbig2012}
have been noted in twisted bilayer TMD \cite{Li2021b,zhao2022realization,foutty2023mapping}, prospects for fractional QSH effect in moir\'e systems remain \textit{terra incognita.} 
}

{In this Letter, we propose a mechanism to realize TRS topological
$\mathbb{Z}_2$ moiré flatbands and establish them as potential platforms to achieve fractional quantum spin Hall effect \cite{PhysRevLett.96.106802,levin2009fractional,PhysRevB.84.165107, PhysRevB.84.165138} through electron interactions in moir\'e heterostructures. 
Our departure is a 
continuum model of small angle twisted bilayer heterostructures which can be applied to bilayer TMD or the paradigmatic Kane-Mele (KM) model\cite{PhysRevLett.95.226801}.
 The KM model was recently realized in a monoelemental honeycomb material--germanene~\cite{PhysRevLett.130.196401}. 
{Two layers of KM model are generally expected to be topologically trivial given the instability of the double pairs of helical edge modes~\cite{PhysRevLett.95.226801}. However, }
our analysis of the 
{small-angle} twisted bilayer KM model identifies topological phase transitions which can be  tuned by the twist angle, interlayer coupling strength and sublattice potential.
The resulting quasi-flat moir\'e bands are characterized by a TRS $\mathbb{Z}_2$ topological invariant signaling a {new} pair of helical edge states. Remarkably, in the 
chiral limit where the $AA$ interlayer coupling vanishes, the wavefunction for each flatband behave as a Kramer's pair of time-reversal invariant LLL wavefunctions containing both holomorphic and anti-holomorphic coordinates related by time-reversal. As such, our work identifies key aspects enabling TRS electron fractionalization in twisted moir\'e bilayers. In particular, we propose classes of many-body wavefunctions hosting fractional QSH stabilized which ground states of short-range interactions.}
{This work thus puts forth a promising route to 
explore fractional QSH in $\mathbb{Z}_2$ flatbands of moir\'e heterostructures. Our consideration may also apply to 
cold atom platforms for which moir\'e engineering has also been made possible recently\cite{Meng2023}.


{\it Model.}-- 
{The low energy description of} both monolayer TMD and germanene with spin $s$ and valley $\tau$ rotated by an angle $\theta$ can be given by
\begin{equation}
  h_{s\tau}(\theta,\bk)=\tau|\bk| v_F \begin{pmatrix}
    0 & e^{-i\tau(\theta_k-\theta)}\\
    e^{i\tau(\theta_k-\theta)} & 0 \\
  \end{pmatrix}+\delta\sigma_z +\lambda s\tau\gamma.\label{eq:hst}
\end{equation}
Here we have set $\hbar=1$. $\delta$ in the case of TMD denotes the sublattice potential difference while in germanene it can arise from coupling to substrate\footnote{Here we approximately assume that $\delta$ is the same for both layers.}. $\lambda$ in both cases stands for the spin-orbit coupling (SOC) strength, and $\gamma=\text{diag}(\gamma_1,\gamma_2)$.  For TMDs\cite{PhysRevLett.108.196802} we have $\gamma=(1-\sigma_z)/2$, while for germanene $\gamma=\sigma_z$ as in the Kane-Mele model\cite{PhysRevLett.95.226801}. 
Eq.\eqref{eq:hst} preserves time-reversal ($\mathcal{T}$) symmetry,
{ and an {\it emergent} U(1) symmetry for the spin $s_z$ component.}
When two layers of TMD or germanene described by Eq.\eqref{eq:hst} are stacked and twisted by a small angle, the moiré pattern develops as is shown in Fig.\ref{fig:band1}(a). The emergent moiré periodicity gives rise to much smaller moiré Brillouin zone shown in Fig.\ref{fig:band1} (b).
{Following \cite{BM},} the continuum model for both twisted bilayer TMD and germanene systems can thus be described in a uniform way.  For spin $s$ and valley $\tau$, the Hamiltonian written explicitly in the two layer space is 
\begin{equation}
  H_{s\tau}= \begin{pmatrix}
    h_{s\tau}(\frac{\theta}{2},\bm{\nabla}) & T_\tau(\br)\\
    T_\tau^\dagger(\br) & h_{s\tau}(-\frac{\theta}{2},\bm{\nabla})
  \end{pmatrix},\label{eq:Hr}
\end{equation}
where $h_{s\tau}(\theta/2,\bm{\nabla})=-i\tau v_F R[\bm{\sigma}\cdot\bm{\nabla}]R^{-1}+\delta\sigma_z+\lambda s\tau\gamma$ (with $R=e^{-i\frac{\theta}{4}\sigma_z}$) is the real space representation of Eq.\eqref{eq:hst}, 
and the local interlayer coupling $T(\br)$ captures the moiré superlattice, 
The Hamiltonian acts on a spinor  $\psi_{s\tau}=(\psi_{A1},\psi_{B1},\psi_{A2},\psi_{B2})^T$, where $A,B$ and $1,2$ are sublattice and layer indices respectively.
As in TBG the interlayer coupling can be approximated by $T_+(\br)=\sum_{n=1}^3 T_{n} e^{-i\bq_n\cdot \br}$ where $\bq_1=k_\theta(0,-1)$, $\bq_2=k_\theta(\sqrt{3}/2,1/2)$ and $\bq_2=k_\theta(-\sqrt{3}/2,1/2)$ with $k_\theta=2k_D\sin(\theta/2)$ being the moiré Brillouin zone length scale. Note that for the other valley $T_-(\br)=T_+^*(\br)$. The three coefficients $T_n$ are $T_{n+1}=w_{AA}I+ w_{AA}'\sigma_z +w_{AB}(\sigma_x \cos \frac{2n\pi}{3} + \sigma_y \sin \frac{2n\pi}{3})$,
with $w_{AA}(w_{AA}')$ and $w_{AB}$ the interlayer tunneling strength in AA- and AB-stacked areas respectively, and $w_{AA}'$ vanishes for germanene but remains nonzero for TMD. 

\begin{figure}
  \includegraphics[width=8.5cm]{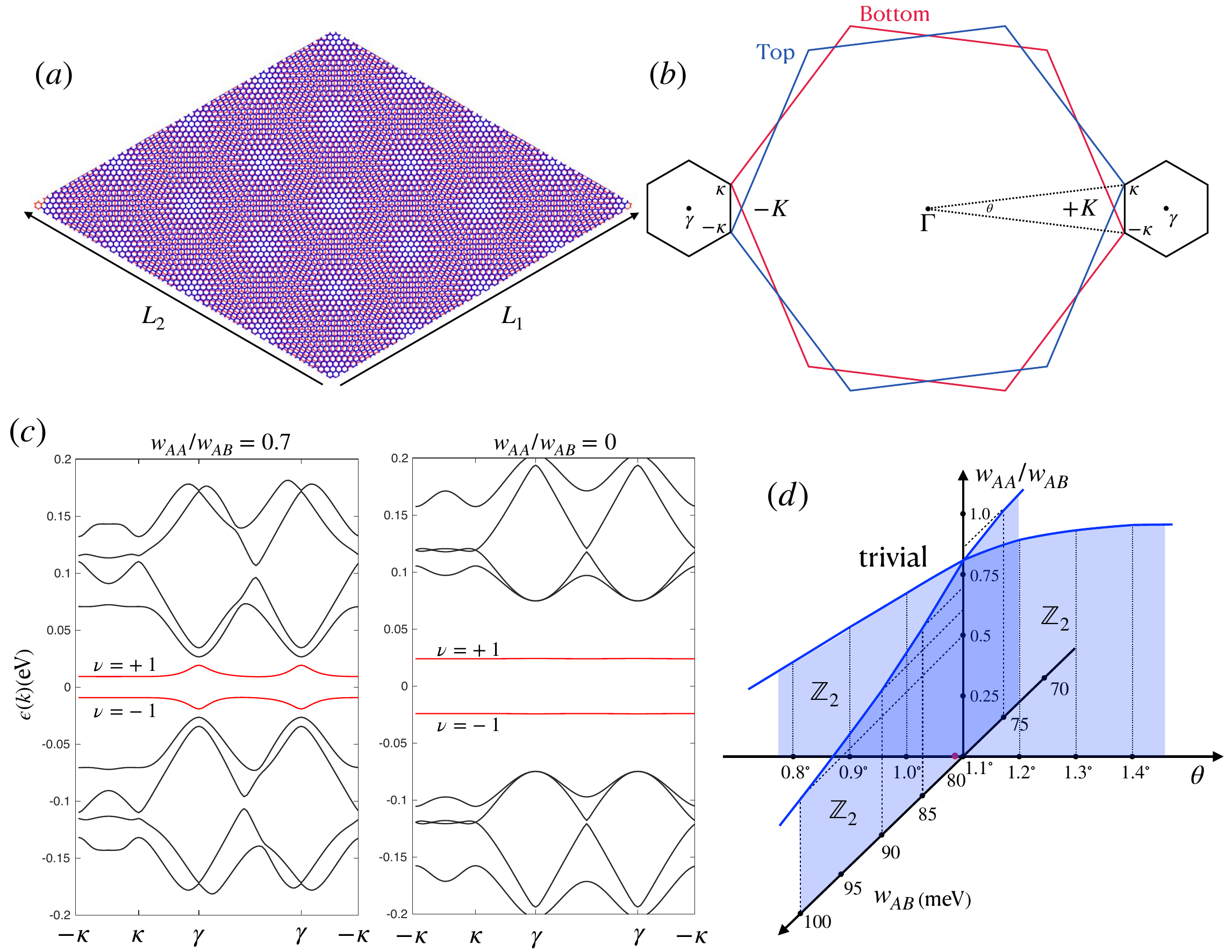}
  \caption{(a)Moiré pattern with system size given by $L_1$ and $L_2$. (b) Moiré Brillouin zone at $\pm K$ valleys. The definition of $\pm\kappa$ is chosen in such a way that $\kappa$ at $+K$ is related to $-\kappa$ at $-K$ valley by time-reversal. (c)Moiré band structure for $+K$ valley at the first magic angle $\theta=1.09^\circ$, obtained by choosing $w_{AB}=80$meV and $\delta=0$. (d) Phase diagram with $\delta\neq0$. For a fixed $w_{AA}$ and $w_{AB}$, tuning the twist angle can result in a topological phase transition from a $\mathbb{Z}_2$ band to a trivial band.}\label{fig:band1}
\end{figure}

{\it $\mathbb{Z}_2$ flatbands.}-- Here we take twisted bilayer germanene as an example. The corresponding moiré band structure is shown in Fig.\ref{fig:band1} (c), which was obtained by setting 
$v_F=5.6\times 10^5 m/s$, and $\lambda=24meV$ \cite{Roome2014,PhysRevLett.130.196401}. 
 Note $v_F$ is around $70\%$ of that in graphene.
Assuming the interlayer coupling is also around $70\%$ of that in TBG, we approximately take $w_{AB}=80meV$ and set $r=w_{AA}/w_{AB}$ as a variable.
The two bands marked in red are the moiré flatbands, separated by a gap due to $\lambda_{\text{so}}\neq0$. The first magic angle at which the moiré bandwidth gets minimized is determined by the condition $\alpha:=w_{AB}/(v_Fk_\theta)\approx0.586$\cite{Tarnopolsky2019}. 
Furthermore, for a given $s\tau$ configuration, the flatbands have nonzero Chern numbers given by $\mathcal{C}=\pm\sgn(s)$, where `$+$' is for the upper bands and `$-$' is for the lower bands.
This gives rise to nontrival topology with TRS which is charaterized by
the $\mathbb{Z}_2$ topological invariant \cite{PhysRevLett.97.036808,RevModPhys.82.3045}
\begin{equation}
  \nu_{\pm}=\frac{\mathcal{C}_{\uparrow,\pm}-\mathcal{C}_{\downarrow,\mp}}{2}\text{mod}2\label{eq:z2}
\end{equation}
Thus the upper moiré band has $\nu_{\pm}=1$ while the lower moiré band has $\nu_{\pm}=-1$. The topological phase can be tuned by $\theta$, $w_{AB}$ and $w_{AA}$, as shown in Fig.\ref{fig:band1}(d). We also confirmed that at even larger $\theta$ (not shown) the system becomes trivial as well, consistent with the $\mathbb{Z}_2$ classification of two layer KM model. 

The band structure for twisted bilayer TMD are similar to those shown in Fig.\ref{fig:band1}(c), except that $\delta$ and $\lambda$ are much larger than those in germanene, and $w_{AA}'$ can be nonzero due to the difference between valence and conduction bands. The large band gap resulting from $\delta$ in TMD has a remarkable consequence of large sublattice polarization, which we will explain in the following. 

{\it Chiral limit.}--
The Hamiltonian in Eq.\eqref{eq:Hr} has an emergent chiral symmetry when $w_{AA}$ and $w_{AA}'$ vanish. 
To see this, 
note one can always find a rotation in spinor space that turns $h_{s\tau}$ onto $x$-$y$ plane while keeping $T_{\tau}(\br)$ off-diagonal. Next by reshuffling the basis $H_{s\tau}$ can be written into an explicitly chiral form (block-off-diagonal).
In this chiral limit and at magic twist angle, 
there are two exactly flat bands for each spin and valley [see Fig.\ref{fig:band1}(c)] of which the energy is determined solely by $\lambda$ and $\delta$. To see this, we can first rotate the basis to $\tilde\chi_{s\tau}=\text{diag}(e^{i\theta\sigma_z/4},e^{-i\theta\sigma_z/4})\psi_{s\tau}\equiv(\chi_{A1},\chi_{B1},\chi_{A2},\chi_{B2})^T $, and then rewrite the Hamiltonian in a new basis
$\chi_{s\tau}=(\chi_{A1},\chi_{A2},\chi_{B1},\chi_{B2})^T\equiv(\chi_A,\chi_B)^T$. The resulting 
eigenvalue equation becomes
\begin{equation}
  \begin{pmatrix}
    \lambda s\tau \gamma_1+\delta & D_\tau^*(-\br)\\
    D_\tau(\br) & \lambda s\tau\gamma_2-\delta
  \end{pmatrix}\begin{pmatrix}
    \chi_A\\
    \chi_B
  \end{pmatrix}=\varepsilon\begin{pmatrix}
    \chi_A\\
    \chi_B
  \end{pmatrix},\label{eq:eigenH}
\end{equation}
where 
\begin{equation}
  D_+(\br)=\begin{pmatrix}
    -2i v_F \partial_{\bar z} &U_+(\br)\\
    U_+(-\br) & -2i v_F \partial_{\bar z}
  \end{pmatrix}, ~D_-(\br)=D_+^*(\br),
\end{equation}
$U_+(\br)=w_{AB}(e^{-i\bq_1\cdot\br}+e^{i\phi-i\bq_2\cdot\br}+e^{-i\phi-i\bq_3\cdot\br})$ and $\partial_{\bar z}\equiv \frac{\partial}{\partial \bar z}=(\partial_x+i\partial_y)/2$.
Eq.\eqref{eq:eigenH} has an apparent solution which is in a fully sublattice-polarized form with either $\chi_A$ or $\chi_B$ remains zero for all $\br$. If we assume $\chi_B=0$, and the eigenvalue problem is solved by $\varepsilon=\lambda s\tau\gamma_1+\delta$ and $D_\tau(\br)\chi_A=0$. The latter condition is in fact identical to the zero energy flatband equation for chiral TBG\cite{PhysRevLett.122.106405}. Note the other set of solution with the opposite sublattice polarization is found by assuming $\chi_A=0$, which has the energy $\varepsilon=\lambda s\tau\gamma_2-\delta$.

\begin{figure}
  \includegraphics[width=8.2cm]{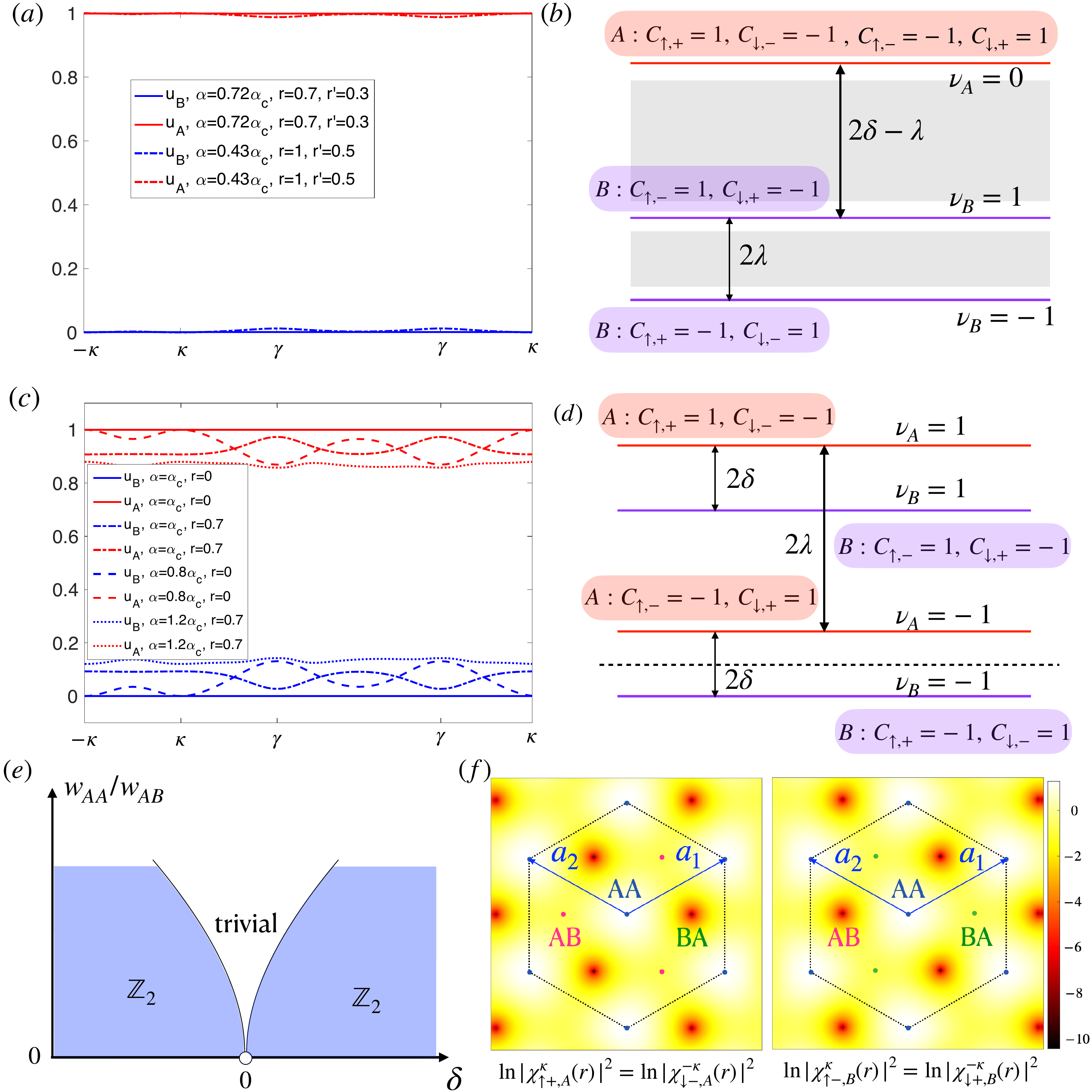}
  \caption{(a-b) Sublattice weight and moiré flat band for twisted bilayer TMD. In (a) we choose the upper flatband with $s=+$ and $\tau=+$ and $\alpha$ is reduced by increasing $\theta$. Due to a large $\delta$ and $\lambda$ in TMD, the sublattice polarization is almost maximized for different $\alpha$, $r$ and $r'$, and between these flatbands, there exist many other bands (shown as gray area). (c-d) Sublattice weight and moiré flat band for twisted bilayer germanene. We again choose $s=+$ and $\tau=+$ for (c). Here the since $\delta$ and $\lambda$ are much smaller than those in TMD, all the four resulting $\mathbb{Z}_2$ flatbands are energetically close to each other.  (e) Schematics of the phase diagram when the Fermi level is at the dashed line shown in (d).  (f)Plots of the wavefunction amplitude at the moiré $\kappa$ point in log-scale. The zeros are located at $\pm(\bm{a}_1-\bm{a}_2)/3$ depending on $s$ and $\tau$.  }\label{fig:chiral}
\end{figure}

The sublattice-polarization is not just a fine tuning effect of the chiral limit and a similar effect has been discussed in lattice models \cite{PhysRevLett.131.026601,PhysRevB.108.L081117,PhysRevB.86.121105}.
In Fig.\ref{fig:chiral} (a) and (c), we show the sublattice weight for the topmost flat band for twisted bilayer TMD and germanene respectively
\footnote{The sublattice weight for the $n$-th band is defined as $u^{(n)}_{A/B}=\sum_{I}|u^{(A/B)}_{I,n}|^2$ where $u^{(A/B)}_{I,n}$ is the eigenstate for the $n$-th band projected to $A/B$ sublattice and $I$ is the shorthand of the indices other than sublattice.}. For TMD bilayers, the sublattice polarization is always almost maximized for a large range of parameters, while for germanene bilayers, it is maximized only when the system is near to chiral limit and magic angle ($\alpha\to\alpha_c$ and $r\to0$).
In Fig.\ref{fig:chiral} (b) and (d) we show the Chern numbers 
for the four flat bands, which help us to further identify the $\mathbb{Z}_2$ flatbands according to Eq.\eqref{eq:z2}. For TMD bilayers, since we neglected the SOC in the conduction band, there are only two $\mathbb{Z}_2$ flatbands with B-sublattice polarization, which are separated by many other bands due to large $\lambda$ [gray area in Fig.\ref{fig:chiral} (b)]. In contrast, the germanene bilayers can host four energetically close $\mathbb{Z}_2$ flatbands, 
If the Fermi level is located in the middle of the lower two bands, then tuning $\delta$ results in a topological phase transition as indicated in Fig.\ref{fig:chiral} (e), where in the chiral limit the trivial regime shrinks to a point at $\delta=0$.

Similar to TBG \cite{PhysRevLett.122.106405}, the wave function for these chiral flat bands can be obtained. 
For simplicity we consider the topmost band. 
For $\tau=+1$, the equation $D_+(\br)\chi_A(\br)=0$ has a $C_3$-rotation symmetry protected solution at the moiré Brillouin corner $\pm\kappa$, which we denote as $\chi_{\uparrow+}^{\pm\kappa}$ (the wavefunction at $\kappa$ and $-\kappa$ are related by $C_3$ rotation so below we use $\chi_{\uparrow+}^{\kappa}$ only). Because $D_+(\br)$ contains only the anti-holomorphic derivative $\partial_{\bar z}$, the general solution can be written as $\chi_{\uparrow+,\bk}(\br)=f_{\bk}(z)\chi_{\uparrow+}^\kappa(\br)$ where $\partial_{\bar z}f_{\bk}(z)=0$. The function $f_{\bk}(z)$ 
must have simple poles to preserve the moiré lattice translation symmetry. But remarkably each component of $\chi_{\uparrow+}^\kappa(\br)$ has zeros at some particular $\br_0$ right at the magic twist angle. Therefore, one can locate the poles of $f_{\bk}(z)$ at these $\br_0$ to make $\chi_{\uparrow+,\bk}(\br)$ a bounded function.
In Fig.\ref{fig:chiral} (f) we plot the norm of the two-component wavefunction $\chi_{s\tau}^{\kappa}(\br)$ on a logarithmic scale. The zeros are located at either $AB$- or $BA$-stacking points, depending on $s$, $\tau$ and sublattice. 
For A-sublattice polarized flat bands, $\br_{0}=(\bm{a}_1-\bm{a}_2)/3$ when measured from the AA-stacking center.

The total wavefunction can be more conveniently expressed when the spatial origin is shifted to $\br_0$, so that
\begin{equation}
  \chi_{\uparrow+,\bk}(\br)=F_{\uparrow+}(\br)N(\bk)\left[e^{i\frac{k^*z}{2}}\sigma\left(z+i\frac{S}{2\pi}k\right)\right]\label{eq:chi1}
\end{equation}
where 
\begin{equation}
  F_{\uparrow+}(\br)=\frac{\chi_{\uparrow+}^\kappa(\br+\br_0)e^{-\frac{\pi a_1^*z^2}{2Sa_1}}}{\vartheta_1(\frac{z}{a_1},\frac{a_2}{a_1})},~N(\bk)=\frac{\pi\vartheta'(0,\frac{a_2}{a_1})}{a_1}e^{\frac{Sa_1^*k^2}{8\pi a_1}}.
\end{equation}
Here the plain forms of $z$, $k$, {\it etc.} are the complex version of the corresponding vectors, for instance $k\equiv k_x+ik_y$, and $S=|\bm{a}_1\times\bm{a}_2|$ is the area of the moiré unit cell. 
$\theta_1(u,\eta)$ is one of the Jacobi theta functions $\vartheta_1(u,\eta)$\cite{KHARCHEV201519} which has the double-periodic properties $\vartheta_1(u+n,\eta)=(-1)^n\vartheta_1(u,\eta)$ and $\vartheta_1(u+n\eta,\eta)=(-1)^ne^{-i\pi(2nu+n^2\eta)}\vartheta_1(u,\eta)$ for $n\in\mathbb{Z}$, and vanishes at $u=n+m\eta$ for $n,m\in\mathbb{Z}$ such that $1/\vartheta_1(u,\eta)$ has simple poles at these positions. As a result, $F(\br)$ becomes regular. The universal part, $[...]$ in the above equation, contains a modified Weierstrass sigma function \cite{10.1063/1.5042618}, which has zeros on the moiré lattice sites, and is related to $\vartheta_1$ via $\sigma(z)=\frac{a_1}{\pi} \exp(\frac{\pi a_1^* z^2}{2S a_1}){\vartheta_1(\frac{z}{a_1},\eta)}/{\vartheta_1'(0,\eta)}$. 
The time-reversal counterpart $\chi_{\downarrow-,\bk}(\br)$ is constructed in the same manner. 
We first notice that the zeros of $\chi_{\uparrow+}^{\kappa}$ and  $\chi_{\downarrow-}^{-\kappa}$ coincide in space [see Fig.\ref{fig:chiral} (f)], and they can be made complex conjugate to each other. Furthermore, the operator $D_-(\br)$ contains only holomorphic derivatives, indicating the wavefunction for $\tau=-1$ valley  contains only anti-holomorphic functions. Clearly this construction is equivalent to taking the complex conjugate of $\chi_{\uparrow+,\bk}$, and we thus have
\begin{equation}
  \chi_{\downarrow-,\bk}(\br)=F_{\downarrow-}(\br)N^*(\bk)\left[e^{-i\frac{kz^*}{2}}\sigma\left(z^*-i\frac{S}{2\pi}k^*\right)\right]\label{eq:chi2}
\end{equation}
with $F_{\downarrow-}(\br)=F_{\uparrow+}^*(\br)$. 
Using the quasi-periodic properties of the theta function, it is straightforward to check that
both of these wavefunctions indeed satisfy Bloch's theorem \footnote{
One caveat here is that  $\chi_{s\tau,\bk}$ is a two-component wavefunction, with each component from distinct layers 
shifted by $\bq_{1}$ from each other. Bloch's theorem for $\chi_{s\tau}$ thus implies $\chi_{s\tau,\bk}(\br+\bm{a})=e^{i\bk\cdot\bm{a}}\text{diag}(1,e^{i\bq_1\cdot\bm{a}})\chi_{s\tau,\bk}(\br)$.}.

Eq.\eqref{eq:chi1} and \eqref{eq:chi2}
are quite similar to the LLL wavefunction on a torus in the symmetric gauge \cite{PhysRevB.31.2529}. The difference is that for the LLL $F(\br)=e^{-|z|^2/4}$ and $N(\bk)=e^{-|k|^2/4}$ (setting magnetic length $l_B=1$ for simplicity), and $\sigma(z)$ is defined for arbitrary lattice as long as the unit-cell area is $S=2\pi$ (one flux quantum for each unit cell). 
In fact, for any 2D ideal flat band with Chern number $\mathcal{C}=1$, its wavefunction can be written in the form of Eq.\eqref{eq:chi1} with some properly chosen $F(\br)$ and $N(\bk)$ \cite{PhysRevLett.127.246403,ledwith2022vortexability}.
The flat band being ideal means that the cell-periodic part, defined as $u_{s\tau,\bk}(\br)=e^{ i \tau\bk\cdot\br}\chi_{s\tau,\bk}(\br)$ is holomorphic in $k$ for spin up and anti-holomorphic for spin down. This follows directly from the identity $\bk\cdot\br=(k^*z+z^*k)/{2}$.
As a key consequence, the quantum geometric tensor
$\eta_{\alpha\beta}(\bk):=\braket{D_\alpha u_{\bk}|D_{\beta} u_{\bk}}$ has vanishing determinant at every $\bk$; here $D_{\alpha}:=\partial_{k_\alpha}-iA_{\alpha}$, $A_{\alpha}=i\braket{u_{\bk}|\partial_{k_\alpha}u_{\bk}}$ being the Berry connection. This in turn implies that the Fubini-Study metric $g_{ab}(\bk)$, the real part of $\eta(\bk),$ is related to the Berry curvature $\bm{\Omega}(\bk)=\nabla_{\bk}\times\bm{A}$  via $g_{\alpha\beta}(\bk)=\frac{1}{2}|\Omega(\bk)|\delta_{\alpha\beta}$. It is known that these properties make the flat band an ideal system to mimic the Girvin-MacDonald-Platzman (GMP) algebra \cite{PhysRevB.33.2481} for the LLL in a strong magnetic field: $[\rho_{\bq_1},\rho_{\bq_2}]=i\Omega\bq_1\times\bq_2\rho_{\bq_1+\bq_2}$ with $\rho_{\bq}$ being the density operator projected to the flat band, if we identify the average Berry curvature $\Omega=\frac{1}{S_{BZ}}\int d\bk \Omega(\bk)$ as the square of the magnetic length $l_B^2$ \cite{PhysRevB.85.241308,PhysRevB.90.165139,PhysRevLett.114.236802}.It is this similarity that makes it possible to obtain quantum (spin) Hall effect in the full or partially filled moiré flat bands. 

{\it Many-body wavefunctions.}-  
Since $\Omega$ plays the same role as $l_B^2$ in the GMP algebra, we can identify each moiré unit cell as the magnetic unit cell which hosts a single magnetic flux. For a parallelogram system with the widths $\bm{L}_1=N_1 \bm{a}_1$ and $\bm{L}_2=N_2\bm{a}_2$  as shown in Fig.\ref{fig:band1}(a), the general twisted periodic boundary condition for each particle on the many-body wavefunction implies $\Psi(\{z_i\}|z_j=L_{1,2})=e^{i\phi_{1,2}}\Psi(\{z_i\}|z_j=0)$, where $\phi_{1,2}$ are not necessarily zero. Following the logic of Ref.\cite{PhysRevB.31.2529, PhysRevResearch.2.023237}, we have for spin-up fermions
\begin{equation}
\label{eq:up spin wf}
  \begin{aligned}
    \Psi_{\uparrow,m}(\{z_j\})=&e^{iKZ}\prod_{j=1}^{N_e}F_{\uparrow+}(\br_j)\prod_{\nu=1}^m \sigma_L(Z-iZ_\nu)\times\\
    &\prod_{i<j}\left[\sigma_L(z_i-z_j)\right]^m
  \end{aligned}
\end{equation}
where we have assumed there are $N_e$ spin-up fermions so that the filling fraction is $1/m=N_e/N_s$. Note that we need to keep $m$ an odd integer in order to maintain fermionic properties. Here $Z=\sum_j z_j$ and the values of $K$ and $Z_0:=\sum_\nu Z_\nu$ are chosen to satisfy 
\begin{equation}
  e^{iKL_{1,2}} = (-1)^{N_s+N_{1,2}}e^{i\frac{\pi L_{1,2}^* Z_0}{N_s S}+i\phi_{1,2}}.\label{eq:BC}
\end{equation}
The sigma function with subscript `L' is defined similarly to the previous discussion, but with the unit cell enlarged to the whole sample spanned by $L_1$ and $L_2$ instead of $a_1$ and $a_2$. Since the sigma function vanishes as $\sigma_L(z)\sim z$ when $z\to0$, this wavefunction scales as $(z_i-z_j)^m$ whenever there are two particles approaching each other, so it can be stablized by some pseudopotential similar to Haldane's.
The difference from Haldane's pseudopotential is that the ideal flat band pseudopotential not only depends on the relative angular momentum between two particles, but also on their center-of-mass (COM), so that the general form of the interactions can be written as $V(\br_1,\br_2)=\sum_{M,m}v_{M,m}\hat{P}_{M,m}$ where $\hat{P}_{M,m}$ is the projector \cite{PhysRevLett.127.246403}. This can be traced back to the fact that the LLL wavefunction obeys the magnetic translation group \cite{PhysRev.134.A1602} while the ideal flatband wavefunction in our case does not. One simple realization that stablize the wavefunction in Eq.\eqref{eq:up spin wf} is to consider sufficiently short range (Hubbard-like) interactions, for which the COM gets frozen, and the pseudopotential can be modeled by $V(\br)=\sum_{m'=0}^{m'<m}v_{m'} (\nabla^2)^{m'}\sum_i\delta(\br-\bm{R}_i)$ where $\bm{R}_i$ denote all lattice sites and all $m'>0$ should be odd.  For the spin-down fermions, the construction is exactly the same,
\begin{equation}
\label{eq:down spin wf}
  \begin{aligned}
    \Psi_{\downarrow,m'}(\{\bar w_j\})=&e^{-i\bar Q\bar W}\prod_{j=1}^{N_e'}F_{\downarrow-}(\br_j)\prod_{\nu=1}^{m'} \sigma_L(\bar W-i\bar W_\nu)\times\\
    &\prod_{i<j}\left[\sigma_L(\bar w_i-\bar w_j)\right]^{m'}
  \end{aligned}
\end{equation}
with $\bar W=\sum_j \bar w_j$, and $\bar Q$ and $\bar W_0:=\sum_\nu \bar W_\nu$ satisfying conditions similar to Eq.\eqref{eq:BC}. 

{Upon combining the two spin components \eqref{eq:up spin wf} and \eqref{eq:down spin wf}, we identify}
\begin{equation}
  \Psi_{\text{FQSH}}(\{z_j,w_j\})=\Psi_{\uparrow,m}(\{z_j\})\Psi_{\downarrow,m}(\{\bar w_j\})\label{eq:FQSH}
\end{equation}
{as a candidate wavefunction describing a  FQSH state with spin filling fraction $\nu_{spin} = \frac{1}{m}$, which hosts conserved spin current and fractional {electronic} excitations \cite{PhysRevLett.96.106802,levin2009fractional,PhysRevB.84.165107, PhysRevB.84.165138}. 
{In short samples compared with the electron mean free path, the presence of a helical edge state may be revealed by the low-temperature quantization of the longitudinal conductance $G = 2 e^2/h$ \cite{konig2007quantum}, expected to persist for an interacting Luttinger liquid edge \cite{Maslov_Landauer1995}. Furthermore, charge fractionalization could be sensed via shot noise measurements \cite{De-Picciotto1998direct,saminadayar1997observation}, providing two complementing experimental signatures of time-reversal symmetric fractionalization.
}

Remarkably, the ideal flat band condition \cite{PhysRevResearch.2.023237,PhysRevLett.127.246403} ensures that \eqref{eq:FQSH} is the ground state of a local time-reversal symmetric Haldane pseudo-potential interaction, {which establishes a microscopic mechanism for}
time-reversal invariant fractional topological order in moir\'e flat band systems. The many-body wavefunction \eqref{eq:FQSH} can be generalized by multiplication by terms $\sim \prod_{r<s}(z_r-\bar w_s)^n$, which represent correlations between opposite spins\cite{PhysRevLett.96.106802, PhysRevB.84.165138}. Achieving such FQSH states would require different local interactions, a question that merits further investigation.}


{\it Discussion.}- We have shown that twisted bilayer 2D materials with spin-orbit coupling (TMD and germanene)
can give rise to ideal flat bands with TRS at magic twist angle and in the chiral limit, serving as an ideal platform for realizing FQSH effect. There are, however, two obstacles that can potentially destroy the FQSH state. The first one is the competition with other symmetry breaking phases, in particular ferromagnetism\cite{mai202314}. The true ground state depends on the details of the interactions, so given that moiré systems have much higher tunability of interactions compared to other platforms, we expect that the FQSH state considered here is indeed in a physically accessible regime. 
The second obstacle is that a finite $w_{AA}$ spoils the ideal flat band condition and renders the Berry curvature more inhomogeneous in $\bk$-space. Comparing energies of different competing states in this case, e.g. using exact diagonalization {and DMRG}, is an interesting question which we leave for a future study.  


{\it Acknowledgements.}- We thank Fengcheng Wu, Trithep Devakul, Ben Feldman, Qiong Ma, Sri Raghu and Steven Kivelson for useful discussions. This research was supported by the Gordon and Betty Moore Foundation’s EPiQS Initiative through GBMF8686 (Y.-M. W), and by the U.S. Department of Energy, Office of Science, Basic Energy Sciences, under 
Award DE-SC0023327 (D. S. and L. H. S.).


\bibliography{tBG}

\newpage

\begin{widetext}
  \begin{center}
  \textbf{\large  ONLINE SUPPLEMENTARY MATERIAL FOR
  \\[.2cm] Time-reversal invariant topological moiré flatband and fractional quantum spin Hall effect}
\\[0.2cm]
Yi-Ming Wu,$^{1}$, Daniel Shaffer$^{2}$, Zhengzhi Wu$^{3}$ and Luiz H. Santos$^{1}$
\\[0.2cm]
{\small \it $^{1}$ Stanford Institute for Theoretical Physics, Stanford University, Stanford, California 94305, USA}

{\small \it $^{2}$ Department of Physics, Emory University, 400 Dowman Drive, Atlanta, GA 30322, USA}

{\small \it $^{3}$ Institute for Advanced Study, Tsinghua University, Beijing 100084, China}

  \vspace{0.4cm}
  \parbox{0.85\textwidth}{In this Supplemental Material we show i) the construction of LLL wave function on a torus using both theta function and sigma function, ii) the construction of many body Laughlin state on a torus and iii) similar construction for the moiré flatband.} 
\end{center}
\end{widetext}

\section{A.LLL wavefunction on a torus and the construction of Laughlin wavefunction.}
The LLL wavefunction on a torus was first studied in Ref.\cite{PhysRevB.31.2529}, where the theta function was used to account for the double-periodicity nature of the wavefunction. There the theta functions is periodic in terms of the boundaries $L_1$ and $L_2$, which necessarily involves a product of many theta functions. Alternatively, one can define the problem on a lattice, and the unit cell is chosen arbitrarily but must enclose an area of $2\pi l_B^2$ where a unit flux passes. The introducing of lattice makes it easy to compare with real lattice system with a flat Chern band. Below we discuss both of these two pictures separately. 

\subsection{LLL without lattice} 
\label{sub:lll_without_lattice}

\subsubsection{single-particle wavefunction}

Here we closely follow the logic of Ref.\cite{PhysRevB.31.2529}. The single particle LLL wavefunction in Landau gauge $\bm{A}(\br)=(-yB,0,0)$ can be written as 
\begin{equation}
  \psi(x,y)=e^{-{y^2}/{2}} f(z).
\end{equation}
Here we have set $l_B^2=1$, and $f(z)$ is a holomorphic function defined on the whole plane. Due to the presence of the exponentially decaying prefactor $e^{-{y^2}/{2}}$, $f(z)$ can be unbounded in the $y$-direction, so the total wavefunction $\psi(x,y)$ can still be normalizable. 
The magnetic translation operator $t(\bm{L})$ acting on a wavefunction is (assuming $\bm{B}$ is in the $z$-direction)
\begin{equation}
  t(\bm{L})\psi(\br)\equiv\psi(\br+\bm{L})=e^{i\bm{L}\cdot[-i\bm{\nabla}-e\bm{A}(\br)]+ie\bm{B}\cdot(\br\times\bm{L})}\psi(\br).
\end{equation}
Suppose we defined the system size as a parallelogram with width $|\bm{L}_1|$ and $|\bm{L}_2|$ and and angle $\varphi$ between them. Then the total number of flux piercing this sample is given by
\begin{equation}
    N_s=\frac{A}{2\pi l_B^2}=\frac{|\bm{L}_1||\bm{L}_2|\sin\varphi}{2\pi}.
  \end{equation}  
  The twisted boundary condition on the wave function implies,
\begin{equation}
  \psi(\bm{L}_{1,2})=\psi(0)e^{i\phi_{1,2}}.\label{eq:pbc}
\end{equation}
Let's further assume $\bm{L}_1$ is in $x$-direction without loss of generality. Then we can immediately see 
\begin{equation}
  \boxed{f(L_1)=f(0)e^{i\phi_1}}.\label{eq:bc1}
\end{equation}
The boundary condition in $\bm{L}_2$ direction also applies to the prefactor $e^{-{y^2}/{2}}$, so we will have
\begin{equation}
  e^{-(|\bm{L}_2|\sin\varphi)^2/2}f(L_2)=f(0)e^{i\phi_2}.
\end{equation}
Using the definition of $N_s$, we can also write the above expression as
\begin{equation}
  \boxed{e^{-\pi N_s \frac{|\bm{L}_2|}{|\bm{L}_1|}\sin\varphi}f(L_2)=f(0)e^{i\phi_2}}.\label{eq:bc2}
\end{equation}
The solution of the above two boxed equations are given by the Jacobi theta function,
\begin{equation}
  f(z)=e^{ik z}\prod_{\nu=1}^{N_s}\vartheta_1\left(\frac{z-z_\nu}{L_1},\tilde\eta\right)\label{eq:productTheta}
\end{equation}
where $\tilde\eta=L_2/L_1$ and the theta function $\vartheta_1$ is defined as 
\begin{equation}
  \vartheta_1(u,\tilde\eta):=-i\sum_{l\in\mathbb{Z}}(-1)^l q^{(l+1/2)^2}e^{i\pi(2l+1)u}, ~q:=e^{i\pi\tilde\eta},
\end{equation}
 which has the following properties,
\begin{equation}
  \begin{aligned}
    \vartheta_1(u+n,\tilde\eta)&=(-1)^n\vartheta_1(u,\tilde\eta),\\
    \vartheta_1(u+n\tilde\eta,\tilde\eta)&=(-1)^ne^{-i\pi(2nu+n^2\tilde\eta)}\vartheta_1(u,\tilde\eta),\\
    \vartheta_1(-u,\tilde\eta)&=-\vartheta_1(u,\tilde\eta),\\
    \vartheta_1(u,\tilde\eta)&=0 ~~~ \text{for}~~ u=n+m\tilde\eta,~m,n\in\mathbb{Z}\\
    \vartheta_1(u,\tilde\eta)&\sim u ~~~ \text{for}~~ u\to0.
  \end{aligned}
\end{equation}

We now need to choose proper $k$ and $z_\nu$ in order to satisfy the boxed boundary condition in Eq.\eqref{eq:bc1} and \eqref{eq:bc2}. Using the properties of $\vartheta_1$ listed above, and defining $z_0=\sum_\nu^{N_s}z_\nu$, it is easy to see that $k$ and $z_0$ satisfy the following equations,
\begin{equation}
  \begin{aligned}
    e^{ikL_1}&=(-1)^{N_s}e^{i\phi_1},\\
    e^{ikL_2}&=(-1)^{N_s}e^{i\left(\phi_2-2\pi z_0/L_1+\pi N_s \frac{|\bm{L}_2|}{|\bm{L}_1|}\cos\varphi\right)}.\label{eq:solutionkz}
  \end{aligned}
\end{equation}
Note the second equation is slightly different from that in Ref.\cite{PhysRevB.31.2529}, and by setting $\varphi=\pi/2$, i.e. when the $\bm{L}_1$ and $\bm{L}_2$ are perpendicular to each other, the equation is reduced to the one in Ref.\cite{PhysRevB.31.2529}. From Eq.\eqref{eq:solutionkz} it is easy to see that, if $(k,z_0)$ is the solution, so is $(k+n_1 2\pi/L_1, z_0-n_1L_2+n_2L_1)$ with $n_1,n_2\in\mathbb{Z}$. The number of the linearly independent solutions is equal to the number of zeros of $f(z)$, which is $N_s$.

\subsubsection{Laughlin wavefunction} 


The construction of the many-body Laughlin wavefunction proceeds as follows. Suppose there are $N_e$ electrons, so the filling factor is given by $N_e/N_s=1/m$. We will assume $m\geq3$ is some odd integer. The wavefunction is given by the ansatz,
\begin{equation}
  \Psi(\{z_i\})=F(Z)\prod_{i<j}g(z_i-z_j)\label{eq:manyb}
\end{equation}
where $F(Z)$ is a function which depends on the center-of-mass coordinate $Z$ and $g(z_i-z_j)$ is the Jastrow factor which only depends on the relative coordinate. Recall the usual Laughlin wavefunction is 
\begin{equation}
  \Psi_{\text{LW}}(\{z_i\})=e^{-\sum_i^{N_e}|z_i|^2/4}\prod_{i<j}(z_i-z_j)^m,
\end{equation}
hence we need 
\begin{equation}
  g(z_i-z_j)\sim (z_i-z_j)^m ~~\text{for}~~ z_i\to z_j.
\end{equation}
One possibility is
\begin{equation}
  g(z)=\left[\vartheta_1\left(\frac{z}{L_1},\tilde\eta\right)\right]^m,
\end{equation}
which leads to the following boundary condition for a generic single particle, say $z_1$
\begin{equation}
  \begin{aligned}
    g(L_1-z_j)&=(-1)^mg(-z_j)\\
    g(L_2-z_j)&=(-1)^{m}e^{-im\pi\left(-2\frac{z_j}{L_1}+\frac{L_2}{L_1}\right)}g(-z_j).
  \end{aligned}
\end{equation}
Therefore, for the product of $N_e-1$ particles, we have
\begin{equation}
  \begin{aligned}
    \prod_{j}g(L_1-z_j)&=(-1)^{N_s-m}\prod_{j}g(-z_j)\\
    \prod_{j}g(L_2-z_j)&=(-1)^{N_s-m}e^{i2\pi m\frac{Z}{L_1}-i(N_s-m)\pi\frac{L_2}{L_1}}\prod_{j}g(-z_j)\label{eq:Pig}
  \end{aligned}
\end{equation}
If we require that the total wavefunction satisfies 
\begin{equation}
  \Psi_{\text{LW}}(\{z_j|z_i=L_{1,2}\})=\Psi_{\text{LW}}(\{z_j|z_i=0\})e^{i\phi_{1,2}}
\end{equation}
then the center-of-mass factor $F(Z)$ must satisfy 
\begin{equation}
  \begin{aligned}
    F(Z+L_1)&=F(Z)(-1)^{N_s-m}e^{i\phi_1},\\
    F(Z+L_2)&=F(Z)(-1)^{N_s-m}e^{-i2\pi m\frac{Z}{L_1}+i(N_s-m)\pi\frac{L_2}{L_1}}e^{i\phi_2}.\label{eq:FZ}
  \end{aligned}
\end{equation}
The general solution for $F(Z)$ can be expressed as 
\begin{equation}
  F(Z)=e^{iKZ}\prod_{\nu=1}^m\vartheta_1\left(\frac{Z-Z_\nu}{L_1},\tilde\eta\right)
\end{equation}
Likewise, we need to properly choose $K$ and $Z_0:=\sum_\nu Z_\nu$ to solve Eq.\eqref{eq:FZ}. This puts constraints on $K$ and $Z_0$, namely
\begin{equation}
  \begin{aligned}
    e^{iKL_1}&=(-1)^{N_s}e^{i\phi_1},\\
    e^{iKL_2}&=(-1)^{N_s}e^{-i2\pi \frac{Z_0}{L_1}}e^{iN_s\pi\tilde\eta} e^{i\phi_2}.
  \end{aligned}
\end{equation}

\subsection{Another choice: LLL wave function with lattice} 
\label{sub:another_choice}

\subsubsection{Symmetric gauge}
Here it is useful to switch to symmetric gauge
 where $\bm{A}=(-yB/2,xB/2,0)$, the magnetic translation operator acting on $\psi(\br)$ has a simple form (again assuming $eB=l_B^{-2}=1$) 
\begin{equation}
  t(\bm{L})=e^{i\bm{L}\cdot(-i\bm{\nabla})+\frac{i}{2}(\br\times\bm{L})\cdot\hat{z}}.\label{eq:trans_symm}
\end{equation}
The wavefunction that simultaneously diagonalize the Hamiltonian and this translation operator can be given by the modified Weierstrass sigma function\cite{10.1063/1.5042618}
\begin{equation}
  \sigma(z)=a_1 e^{\frac{\pi a_1^* z^2}{2A a_1}}\frac{\vartheta_1\left(\frac{z}{a_1},\eta\right)}{\vartheta_1'(0,\eta)}
\end{equation}
where $\vartheta_1$ is the theta function defined above, and $\vartheta_1'$ denotes its derivative. $A$ is the area of the unit cell, and in the lattice spanned by $a_1$ and $a_2$ we have $A=2\pi$. Upon translated by a lattice $l=m a_1 + n a_2$ with $m,n\in \mathbb{Z}$, it changes as 
\begin{equation}
  \sigma(z+l)=\xi(l)e^{\frac{\pi l^*}{A}(z+l/2)}\sigma(z)\label{eq:sigmatrans}
\end{equation}
where $\xi(l)=1$ if $l/2$ is also on the lattice and $\xi(l)=-1$ otherwise. In particular, if $l=a_{j}$ with $j=1,2$, we have
\begin{equation}
  \sigma(z+a_{j})=-e^{\frac{a_{j}^*}{2}(z+a_{j}/2)}\sigma(z).\label{eq:sigmatrans2}
\end{equation}
Now if we write the wavefunction as
\begin{equation}
  \boxed{\psi_{S}(x,y)=e^{-|z|^2/4}f_k(z)}
\end{equation}
then the holomorphic function $f_k(z)$ is given by
\begin{equation}
  \boxed{f_k(z)=\sigma(z+ik)e^{-|k|^2/4+ik^*z/2}}.
\end{equation}
An important property of writing the LLL wave function using this modified sigma function is that it does not dependent on the specific choice of the lattice, i.e. it's modular-invariant. One can design an artificial lattice spanned by the lattice vector $\bm{a}_1$ and $\bm{a}_2$ with the unit cell area being $2\pi$ (it is $2\pi l_{B}^2$ when $l_B$ is not set to unity). Then the system under consideration can be described using two integers $N_1$ and $N_2$, such that $L_1=N_1a_1$ and $L_1=N_2a_2$, and the total number of fluxes passing through the system is given by $N_s=N_1N_2$.
Under translation $\bm{l}=m \bm{a}_1+ n \bm{a}_2$, this wavefunction transforms as 
\begin{equation}
  \begin{aligned}
    &\psi_{S}(\br+\bm{l})=e^{-|z+l|^2/4}f_k(z+l)\\
    &=e^{-\frac{|z|^2+|k|^2}{4}-\frac{|l|^2}{4}-\frac{z^*l+l^* z}{4}+i\frac{k^* z+k^* l}{2}}\sigma(z+l+ik)\\
    &=\xi(l)e^{i\frac{k^* l+l^* k}{2}+\frac{l^* z-z^* l}{4}}\psi_{S}(\br)\\
    &=\xi(l)e^{i\bk\cdot\bm{l}+\frac{i}{2}(\br\times\bm{l})\cdot \hat{z}}\psi_{S}(\br).
  \end{aligned}
\end{equation}
Note this wavefunction transforms in a similar but not exactly the same way as Bloch wavefunction transforms under spatial translation. 
Using this properties, one can explicitly show that 
\begin{equation}
  t(\bm{l}_1) t(\bm{l}_2)\psi_S(\br)=t(\bm{l}_2) t(\bm{l}_1)\psi_S(\br) e^{i(\bm{l}_1\times\bm{l}_2)\cdot\hat{z}}
\end{equation}
as it should be. 
The periodic boundary condition in Eq.\eqref{eq:pbc} then implies $\bk$ must obey 
\begin{equation}
  \begin{aligned}
    \bk\cdot\bm{L}_1&=2\pi n_1+N_1\pi+\phi_1,\\
    \bk\cdot\bm{L}_2&=2\pi n_2+N_2\pi+\phi_2,
  \end{aligned}
\end{equation}
with $n_1,n_2\in\mathbb{Z}$.

Using the relation between sigma function and the theta function, it is possible to express the wavefunction only in terms of theta function. In addition, without loss of generality, we can always choose $a_1$ to be real. It is then easy to see
\begin{equation}
  \boxed{\psi_S(x,y)=e^{i \frac{\bk\cdot\bm{a}_1}{a_1}z}\frac{a_1e^{-\frac{|z|^2-z^2}{4}-\frac{|k|^2+k^2}{4}}}{\vartheta_1'(0,\eta)}{\vartheta_1\left(\frac{z+ik}{a_1},\omega\right)}}.
\end{equation}
Note this expression is different from Eq.\eqref{eq:productTheta} in a sense that it contains only one holomorphic theta function. But the number of zeros enclosed by the sample boundary remains the same. By making use of Eq.\eqref{eq:zk}, it is easy to realize that the function 
\begin{equation}
  u_{\bk}:=\psi_S(\br)e^{-i\bk\cdot\br}\equiv N_{\bk}\tilde{u}_k(\br)
\end{equation}
is a normalized $k$-holomorphic function $\tilde{u}_k(\br)$ times a $\bk$-dependent complex normalization factor $N_{\bk}$ (such that $||u_{\bk}||^2=|N_{\bk}|^2$). This observation is useful, since the quantum geometric tensor $\eta(\bk)$, defined as 
\begin{equation}
  \eta_{ab}(\bk)=\frac{\braket{\partial_{k_a}u_{\bk}|\partial_{k_b}u_{\bk}}}{|N_{\bk}|^2}-\frac{\braket{\partial_{k_a}u_{\bk}|u_{\bk}}\braket{u_{\bk}|\partial_{k_b}u_{\bk}}}{|N_{\bk}|^4}\label{eq:gtensor}
\end{equation}
is actually independent of $N_{\bk}$. A simple manipulation shows that when substituting $u_{\bk}=N_{\bk}\tilde{u}_{k}$ into this definition, the derivatives of $N_{\bk}$ from the first and the second part of Eq.\eqref{eq:gtensor} cancel out, which leads to
\begin{equation}
  \eta_{ab}(\bk)={\braket{\partial_{k_a}\tilde u_{k}|\partial_{k_b}\tilde u_{k}}}-{\braket{\partial_{k_a}\tilde u_{k}|\tilde u_{k}}\braket{\tilde u_{k}|\partial_{k_b}\tilde u_{k}}}.\label{eq:gtensor2}
\end{equation}
Therefore, the factor $N_{\bk}$, although depending on $\bk$, does not determine the properties of the ideal flatband. 

We close by noting that the wavefunction in Landau gauge can be readily obtained by applying the gauge transformation, namely,
\begin{equation}
  \begin{aligned}
    \psi_{L}(x,y)&=\psi_S(x,y) e^{-ixy/2}\\
      &=e^{i \frac{\bk\cdot\bm{a}_1}{a_1}z}\frac{a_1e^{-\frac{y^2}{2}-\frac{|k|^2+k^2}{4}}}{\vartheta_1'(0,\eta)}{\vartheta_1\left(\frac{z+ik}{a_1},\omega\right)}.
  \end{aligned}
\end{equation}

\subsubsection{Laughlin wavefunction}

Using sigma function, we can write the ansatz for the many-body wavefunction similar to that in Eq.\eqref{eq:manyb}, but since we will be using sigma functions, we write the ansatz as 
\begin{equation}
  \Psi(\{z_i\})=e^{-\sum_i\frac{|z_i|^2}{4}}F(Z)\prod_{i<j}g(z_i-z_j),
\end{equation}
 where $g(z)$ is now given by
\begin{equation}
  g(z)=[\sigma_L(z)]^m
\end{equation}
and 
\begin{equation}
  \sigma_L(z)=\frac{L_1}{\pi}e^{\frac{L_1^* z^2}{4N_sL_1}}\frac{\vartheta_1\left(\frac{z}{L_1},\tilde\eta\right)}{\vartheta'(0,\tilde\eta)}.
\end{equation}
Similar to Eq.\eqref{eq:sigmatrans2},
\begin{equation}
  \sigma_L(z+L_{1,2})=-e^{\frac{L_{1,2}^*}{2N_s}(z+L_{1,2}/2)}\sigma(z).
\end{equation}
Clearly $g(z)$ still scales as $z^m$ when $z\to0$. Then it is easy to see, similar to Eq.\eqref{eq:Pig}, we now have
\begin{equation}
  \begin{aligned}
    \prod_jg(L_1-z_j)&=(-1)^{N_s-m}e^{\frac{|L_1|^2}{4N_s}(N_s-m)-\frac{mL_1^*}{2N_s}Z}\prod_j g(-z_j),\\
    \prod_jg(L_2-z_j)&=(-1)^{N_s-m}e^{\frac{|L_2|^2}{4N_s}(N_s-m)-\frac{mL_2^*}{2N_s}Z}\prod_j g(-z_j).
  \end{aligned}
\end{equation}
Accordingly, the periodic boundary condition on the many-body wave function, when applied to one of the many particles,  leads to the following constraints for $F(Z)$, namely 
\begin{equation}
  \begin{aligned}
    \frac{F(Z+L_1)}{F(Z)}&=(-1)^{-N_s+m}e^{\frac{mL_1^*}{2N_s}(Z+\frac{L_1}{2})}e^{i\phi_1},\\
    \frac{F(Z+L_2)}{F(Z)}&=(-1)^{-N_s+m}e^{\frac{mL_2^*}{2N_s}(Z+\frac{L_2}{2})}e^{i\phi_2}.
  \end{aligned}
\end{equation}
These equations are solved by assuming  the following general form 
\begin{equation}
  F(Z)=e^{iK Z}\prod_{\nu=1}^m \sigma_L(Z-iZ_\nu).
\end{equation}
Introducing $Z_0=\sum_{\nu=1}^m Z_\nu$, then the parameter $K$ is determined via
\begin{equation}
  \begin{aligned}
    KL_1 &= \frac{L_1^*Z_0}{2N_s}+\pi N_s+\phi_1+2n_1\pi\\
    KL_2 &= \frac{L_2^*Z_0}{2N_s}+\pi N_s+\phi_2+2n_2\pi\\
  \end{aligned}\label{eq:KLZ}
\end{equation}
with $n_1,n_2\in\mathbb{Z}$.
It is also straightforward to see that under translation operation
\begin{equation}
  t(\bm{L}_1) t(\bm{L}_2)\Psi(\{z_i\})=t(\bm{L}_2) t(\bm{L}_1)\Psi(\{z_i\}) e^{i(\bm{L}_1\times\bm{L}_2)\cdot\hat{z}}.
\end{equation}

\section{B. Laughlin wavefunction for a generic ideal flatband} 
\label{sec:c_laughlin_wavefunction_for_a_generic_ideal_flatband}
From above we see the single particle LLL wavefunction (in symmetric gauge) can be conveniently written as
\begin{equation}
  \psi(\br)=e^{-\frac{|z|^2}{4}}e^{-\frac{|k|^2}{4}}\left[e^{i\frac{k^*z}{2}}\sigma\left(z+i\frac{A}{2\pi}k\right)\right],\label{eq:LLLS}
\end{equation}
where we re-introduced $A=2\pi l_B^2$ but in our convention $l_B=1$ it reduces to $A=2\pi$.
It contains three part. The first one is the factor $e^{-\frac{|z|^2}{4}}$ which depends only on $\br$ and makes sure the wavefunction decays at large distances.
The second term $e^{-\frac{|k|^2}{4}}$ is a factor which depends solely on $\bk$. 
The the last term inside $[...]$ has the nice property that it becomes a holomorphic function in $k$ when multiplied by the factor $e^{-i\bk\cdot\br}$, as
\begin{equation}
  \bk\cdot\br = \frac{1}{2}\left(k^* z+ z^* k\right).
\end{equation}
The manybody Laughlin wavefunction is given by
\begin{equation}
  \begin{aligned}
    \Psi(\{z_i\})=&e^{iK Z}\left(\prod_{i=1}^{N_e}e^{-\frac{|z_i|^2}{4}}\right)\prod_{\nu=1}^m \sigma_L(Z-iZ_\nu)\times\\
    &\prod_{i<j}\left[\sigma_L(z_i-z_j)\right]^m,\label{eq:Laughlin}
  \end{aligned}
\end{equation}
with $K$ and $Z_0=\sum_\nu Z_\nu$ satisfying Eq.\eqref{eq:KLZ}.

In fact, as suggested in Ref.\cite{PhysRevLett.127.246403}, any ideal flatband wavefunction with Chern number $\mathcal{C}=1$ can be written in a way similar to Eq.\eqref{eq:LLLS}, namely
\begin{equation}
  \psi_{\text{IFB}}(\br)=F(\br)N(\bk)\left[e^{i\frac{k^*z}{2}}\sigma\left(z+i\frac{A}{2\pi}k\right)\right]\label{eq:IFB}
\end{equation}
where $F(\br)$ depends on the lattice details, and $N(\bk)$ is some normalization factor which is less important as we already see in Eq.\eqref{eq:gtensor} and \eqref{eq:gtensor2}.

As an example, now let's come back to the magic angle chiral limit flatband wavefunctoin for, say $\mathcal{C}=+1$, 
\begin{equation}
  \begin{aligned}
    \chi_{+,\bk}(\br)&=e^{i\bk\cdot \bm{a}_1 \frac{z-z_0}{a_1}}\frac{\vartheta_1\left(\frac{z-z_0}{a_1}-\frac{k}{b_2},\eta\right)}{\vartheta_1\left(\frac{z-z_0}{a_1},\eta\right)}\chi_{+}^\kappa(\br)\\
    &=e^{i\bk\cdot \bm{a}_1 \frac{z-z_0}{a_1}}\frac{\vartheta_1\left(\frac{z-z_0+i\frac{A}{2\pi}k }{a_1},\eta\right)}{\vartheta_1\left(\frac{z-z_0}{a_1},\eta\right)}\chi_{+}^\kappa(\br).
  \end{aligned}
\end{equation}
It is more convenient to work with the wavefunction with origin shifted to $\br_0$, so we define a new $\tilde\chi_{+,\bk}(\br)=\chi_{+,\bk}(\br+\br_0)$. After some manipulation we rewrite it is as 
\begin{equation}
  \begin{aligned}
    \tilde\chi_{+,\bk}(\br)=&\frac{\chi_{+}^\kappa(\br+\br_0)e^{-\frac{\pi a_1^*z^2}{2Aa_1}}}{\vartheta_1\left(\frac{z}{a_1},\eta\right)}\frac{\pi\vartheta'(0,\eta)}{a_1}e^{\frac{Aa_1^*k^2}{8\pi a_1}}\times\\
  &\left[e^{i\frac{k^*z}{2}}\sigma\left(z+i\frac{A}{2\pi}k\right)\right].
  \end{aligned}
\end{equation}
From this expression we can easily read off the factors $F(\br)$ and $N(\br)$ introduced in Eq.\eqref{eq:IFB} in this case.

As a directly generalization of Eq.\eqref{eq:Laughlin}, the manybody wavefunction ansatz can be written by modifying $F(\br)$ accordingly. Therefore, for the magic angle chiral limit flatband, the manybody Laughlin wavefunction ansatz is given by 
\begin{equation}
  \begin{aligned}
    \Psi_{\text{IFB}}(\{z_i\})=&e^{iK Z}\left(\prod_{i=1}^{N_e}\frac{\chi_{+}^\kappa(\br_i+\br_0)e^{-\frac{\pi a_1^*z_i^2}{2Aa_1}}}{\vartheta_1\left(\frac{z_i}{a_1},\eta\right)}\right)\times\\
    &\prod_{\nu=1}^m \sigma_L(Z-iZ_\nu)\prod_{i<j}\left[\sigma_L(z_i-z_j)\right]^m.\label{eq:LaughlinIFB}
  \end{aligned}
\end{equation}
Likewise, by using the quasiperiodic properties of $\sigma_L$ and $\vartheta_1$, it is easy to see the values of $K$ and $Z_0=\sum_\nu Z_\nu$ must satisfy
\begin{equation}
  \begin{aligned}
    e^{iKL_1} &= (-1)^{N_s+N_1}e^{i\frac{\pi L_1^* Z_0}{N_s A}+i\phi_1},\\
    e^{iKL_2} &= (-1)^{N_s+N_2}e^{i\frac{\pi L_2^* Z_0}{N_s A}+i\phi_2},
  \end{aligned}
\end{equation}
in order to meet the periodic boundary conditions,
which are similar to Eq.\eqref{eq:KLZ}. 

\end{document}